\documentclass[12pt]{article}
\usepackage[pdftex]{graphicx}
\usepackage{caption}
\usepackage{wrapfig}
\usepackage{latexsym}
\usepackage[pdftex]{color}
\usepackage{pifont}
\usepackage{epstopdf}
\usepackage{tikz}
\usepackage{mathtools}
\usetikzlibrary{decorations.pathmorphing,patterns}

\newcommand{\bmp}[2][t]{\begin{minipage}[#1]{#2}}
\newcommand{\emp}{\end{minipage}}

\usepackage{amsmath}
\usepackage{amssymb}
\usepackage{textcomp}
\usepackage{tcolorbox}
\usepackage[title]{appendix}

\usepackage{sidecap}

\usepackage[square,numbers,sort&compress]{natbib} 

\newcommand{\bblubox}[1]{\begin{tcolorbox}[colframe=blue!75!white,title=#1]}
\newcommand{\eblubox}{\end{tcolorbox}}

\newtcolorbox{mybox}{colback=teal!10!white,colframe=blue!75!white}
\usepackage{wrapfig} 

\begin{document}

\title{Frequency dependence of prestin:  \\Intrinsic transition rates and viscoelastic relaxation} 
\author{Kuni H Iwasa, NIDCD, NIH\\ Bethesda, Maryland 20892, USA}
\date{Presented:  at Assoc.\ Res.\ Otolaryngol. meeting \\28 January 2020, San Jose, CA\\ 
Reformatted and corrected: 31 January 2020}
\maketitle

\section*{Abstract}
Outer hair cells in the inner ear is important for the sensitivity, frequency selectivity, and dynamic range of the mammalian ear. Such a physiological role of outer hair cells is thought to be based an amplifying effect of those cells, which have mechanosensitive their hair bundles and motile cell body embedding prestin, a membrane protein with electromechanical coupling at high density. Mechanical power production by outer hair cells were previously evaluated based on an assumption that the intrinsic transition rates of this motile protein is faster than the operating frequency.
This report generalizes the previous treatment by incorporating intrinsic transition rates of the motile membrane protein. It was found that the transition rates attenuates mechanical power production of an outer hair cell such that viscous drag cannot be counteracted beyond the characteristic frequency of gating.

\section*{Introduction}
Fast motile activity (\emph{electromotility}) of outer hair cells (OHCs) is based on conformational transitions (gating) of prestin, a molecule with mechanoelectric coupling, with which motile response is elicited by the receptor potential. A charge movement results in extra membrane capacitance (nonlinear capacitance) and area changes in the axial displacement of the cell. 

The speed of conformational transitions, which is critical for the function of OHCs, should depend on both  ``intrinsic'' transition rates and mechanical load \cite{SantosSacchi2019}. 

Generation of mechanical power, which is critical to the physiological function of these cells, has been examined assuming infinitely fast intrinsic transitions \cite{Iwasa2016,Iwasa2017}. What is the effect of finite intrinsic transition rates on power generation?

\subsubsection*{Specific Questions}

\begin{itemize}
\item How to describe both intrinsic transition rates and mechanical load?
\item How much power output is affected by finite ``intrinsic'' transition rates?
\item What is amplifier gain and limiting frequency?
\end{itemize}

\color{black}
\section*{Intrinsic Transition Rates} 
\begin{wrapfigure}{l}{0.2\textwidth} 
\vspace{-24pt} 
\begin{center}
\includegraphics[width=0.18\textwidth]{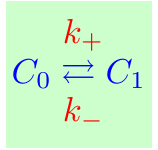} \end{center}
\vspace{-24pt} 
\vspace{1pt}
\caption{}
\label{fig:rates}
\end{wrapfigure}

Consider a membrane molecule with two discrete conformational states C$_0$ and C$_1$ and let the transition rates $k_+$ and $k_-$ (Fig.\ \ref{fig:rates}).

Let $P_1$ the probability that the molecule in state C$_1$.  Then, the probability $P_1$ can be expressed by the transition rates
\begin{align}\label{eq:P1}
\frac{P_1}{1-P_1}=\frac{k_+}{k_-}=\exp[-\beta q(V-V_0)],
\end{align}
where $q$ is the charge transferred across the membrane during conformational changes, $V$ the membrane potential, $V_0$ a constant, and $\beta=1/k_BT$ with the Boltzmann constant and $T$ the temperature. 

If $q>0$, $P_1\Downarrow$  for $V\Uparrow$. For prestin in outer hair cells, if we choose C$_1$ as the shortened state, the unit length change $a<0$ and $q<0$. Length $a$ does not appear in Eq.\ \ref{eq:P1}.

The transition rates that satisfy Eq.\ \ref{eq:P1} can be given by
\begin{subequations}\label{eq:kpm}
\begin{align}
k_+&=\exp[-\alpha\beta q(V-V_0)], \\
k_-&=\exp[(1-\alpha)\beta q(V-V_0)],
\end{align}
\end{subequations}
where $\alpha$ is an arbitrary constant.

The time dependence of $P_1$ can be expressed by the rate equation
\begin{align}
\frac {d}{dt}P_1=k_+-(k_++k_-)P_1.
\end{align}

Introduce a sinusoidal voltage changes of small amplitude $v$ on top of constant voltage $\overline V$, i.e. $V=\overline V+v\exp[i\omega t]$, where $\omega$ is the angular frequency and $i=\sqrt{-1}$. Then the transition rates are time dependent due to the voltage dependence Eq.\ \ref{eq:P1}, satisfying
\begin{align}\label{eq:kratio}
\frac{k_+}{k_-}=\frac{\overline{k}_+}{\overline{k}_-}(1-\beta qv\exp[i\omega t]),
\end{align}
where $\overline{k}_+$ and $\overline{k}_-$ are time independent, and small amplitude $v$ implies $\beta qv \ll 1$. A set of $k_+$ and $k_-$ that satisfies Eq.\ \ref{eq:kratio} can be expressed
\begin{subequations}
\begin{align}\label{eq:kp}
k_+&=\overline{k}_+(1-\alpha \beta qv\exp[i\omega t]),\\ \label{eq:km}
k_-&=\overline{k}_-\{1-(1-\alpha) \beta qv\exp[i\omega t]\}.
\end{align}
\end{subequations}

If we express $P_1=\overline{P}_1+p_1\exp[i\omega t]$, we have respectively for the 0th and 1st order terms \cite{i1997}
\begin{subequations}
\begin{align}
\overline P_1&=\frac{\overline{k}_+}{\overline k_++\overline k_-},\\ \label{eq:p_1}
p_1 &=-\frac{\overline{k}_+\overline{k}_-}{\overline k_++\overline k_-}\cdot\frac{\beta qv}{i\omega+\overline{k}_++\overline{k}_-}.
\end{align}
\end{subequations}
Notice that $p_1$ does not depend on the factor $\alpha$. 

Eq.\ \ref{eq:p_1} leads to voltage-driven complex displacement $ap_1\exp[i\omega t]$ with
\bblubox{}
\begin{align}\label{eq:ap1}
p_1=- \overline{P}_\pm\cdot\frac{\beta qv}{1+i\omega/\omega_g},
\end{align}
\eblubox
\noindent where $ \overline{P}_\pm=\overline{P}_1(1-\overline{P}_1)$.
The amplitude $|x|$ of the motile response is given by
\begin{align}
 |x|^2=\frac{(\beta a q \overline{P}_\pm )^2}{1+(\omega/\omega_g)^2}\cdot v^2.
\end{align}
\begin{wrapfigure}{l}{0.65\textwidth} 
\vspace{-20pt} 
\includegraphics[width=0.6\textwidth]{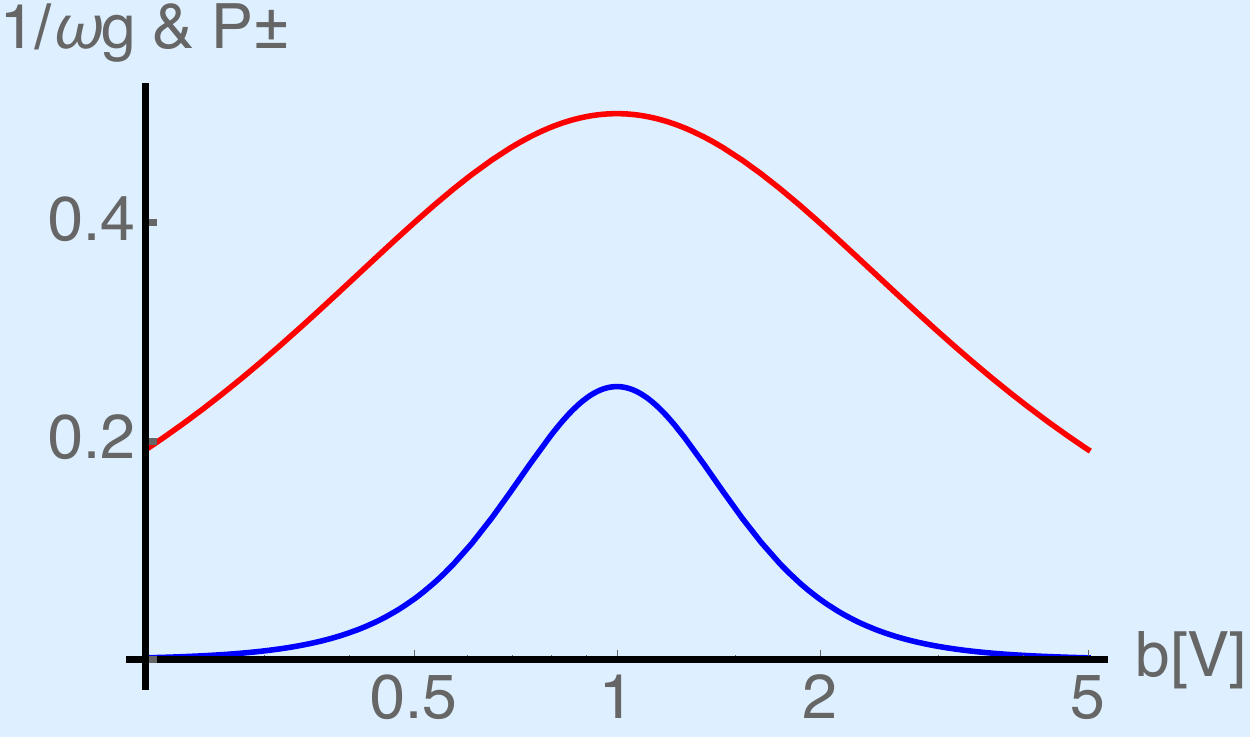}
\caption{\color{red}$1/\omega_g$ \color{black}and \color{blue}$\overline P_\pm$ \color{black}plotted against $b[V]$, a monotonic function of $V$.}
\label{fig:v_dep}
\vspace{-20pt} 
\vspace{1pt}
\end{wrapfigure}

Charge displacement is expressed by $qp_1$, the contribution to complex admittance $Y(\omega)$ is is given by $(q/v)dp_1/dt$. The contribution to the membrane capacitance is $C_\mathrm{nl}(\omega)=Im[Y(\omega)]/\omega$, leading to
\begin{align}
\label{eq:cap_eq}
C_\mathrm{nl}(\omega)=\frac{\beta q^2\overline{P}_\pm}{1+(\omega/\omega_g)^2}.
\end{align}

The roll-off frequency $\omega_g$ due to gating is expressed by
$\omega_g=\overline{k}_++\overline{k}_-$,
which is voltage dependent. Using Eqs.\ \ref{eq:kpm}, we obtain
\begin{align}
\omega_g=\exp[-\alpha\beta q(\overline V-V_0)]+\exp[(1-\alpha)\beta q(\overline V-V_0)].
\end{align}
This means that $1/\omega_g$ rises at both ends of the membrane potential because $\alpha$ can take any value between 0 and 1. That means $\omega_r$ can be asymmetric unless $\alpha=1/2$. In the following we assume $\alpha=1/2$ for the simplicity.

If we define $b(V)=\exp[-\beta q(\overline V-V_0)/2]$, then
\begin{align}
1/\omega_g=1/[b(V)+1/b(V)],
\end{align}
which resembles the voltage dependence of nonlinear capacitance at low frequencies, i.e. $\omega \rightarrow 0$ (See Fig.\ \ref{fig:v_dep}). 
 
\section*{Effect of Mechanical Load} 
If the membrane protein has mechanoelectric coupling, charge transfer is affected by mechanical factors. For simplicity, we approximate a cylindrical cell such as a cochlear outer hair cell with a one-dimensional object.

\begin{wrapfigure}{l}{0.46\textwidth} 
\vspace{5pt}
\vspace{-20pt} 
\includegraphics[width=0.4\textwidth]{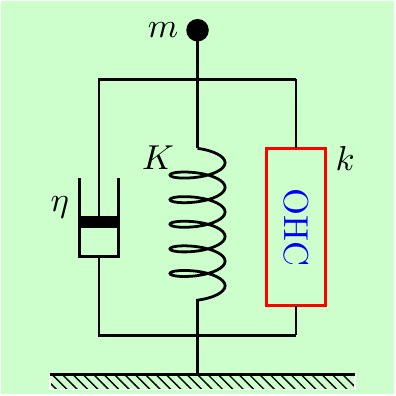}
\caption{Mechanical load imposed on an OHC.}
\label{fig:schem}
\vspace{-20pt} 
\end{wrapfigure}

Suppose charge transfer $q$ is associated with a change $a$ in the length of the cell, Eq.\ \ref{eq:P1} should be replaced by
\begin{align}\label{eq:P1qa}
\frac{P_1}{1-P_1}=\frac{k_+}{k_-}=\exp[-\beta [q(V-V_0)+aF]],
\end{align}
where $F$ is the axial force on the cell. The revised transition rates $k_+$ and $k_-$ also depend on the mechanical factor
\begin{subequations}\label{eq:kpmF}
\begin{align}
k_+&=\exp\left[-\frac \beta 2 [q(V-V_0)+aF]\right], \\
k_-&=\exp\left[+\frac \beta 2[q(V-V_0)+aF]\right].
\end{align}
\end{subequations}
For the rest of the present paper, the dependence on the value of the parameter $\alpha$ is does not appear except for in $\omega_g$.
\section*{Generalization}
A standard way of describing mechanical load is equation of motion. If the system has a mass $m$, and drag coefficient $\eta$ (Fig.\ \ref{fig:schem}), the displacement $X(t)$ follows the equation of motion when it is subjected to an external force $F(t)$
\begin{align}
 \left(m\frac {d^2}{dt^2}+\eta\frac{d}{dt}\right)X(t)=F(t).
\end{align}
It force $F$ is periodic with angular frequency $\omega$, the equation of motion can be expressed as
\begin{align}
( -\omega^2 m+i\eta\omega)x=f(\omega),
\end{align}
where $x$ and $f$ are, respectively, the amplitudes of displacement $X$ and force $F$ with frequency $\omega$.


If the external force $F$ is due to the external load alone, we have $F=-KX$ and $f=-Kx$, which are familiar equations of motion. In such a case equilibrium condition is $X=0$ and $x=0$. In our system, however, movement is driven by a deviation from Boltzmann distribution. If we change the voltage and the displacement simultaneously, the system goes from one equilibrium to another. However, if we change the voltage alone, mechanical displacement must change to establish equilibrium. For this reason, 
\begin{align}\label{eq:ap1v}
p_\infty=-\beta \overline P_\pm \cdot\frac{qv+a^2n\tilde Kp}{1+i\omega/\omega_g}
\end{align}
with $\tilde K=kK/(k+K)$, is the goal of the change, while the current condition is represented by $p$. Note that the quantity $p_\infty$ is similar to $p_1$ in Eq.\ \ref{eq:ap1} in that it satisfies Boltzmann distribution. However, it depends both $v$ and $p$ because the energy term has both electrical and mechanical terms as expressed by Eqs.\ 
\ref{eq:kpmF} in the manner similar to the case of infinitely fast gating \cite{Iwasa2016}.

If the difference between $p_\infty$ and $p$ is small, the driving force can be proportional to the difference $p_\infty-p$. Thus, the driving force can be expressed
\begin{align}
f=k\cdot an(p_\infty-p) 
\end{align}
in the manner similar to the case of fast gating \cite{Iwasa2016}. The presence of external elastic load $K$ makes the displacement $x$ expressed by $x=anp\cdot k/(k+K)$, where $n$ is the number of motile elements. By choosing $p$ as the variable of the equation, we have
\begin{align}\label{eq:eom_fre}
\left[-\omega^2 m+i \eta\omega\right]p&=(k+K)(p_\infty-p),
\end{align}
which has a familiar form for the equation of motion.

By introducing the explicit form of $p_1$, this equation turns into
\begin{mybox}
\begin{align}\label{eq:p1}
  [-(\omega/\omega_r)^2+i \omega/\omega_\eta+1+\delta^2]p=\frac{\beta \overline P_\pm}{1+i\omega/\omega_g}\cdot qv,
\end{align}
\end{mybox}
\noindent where $\delta^2=\beta \overline P_\pm na^2\tilde K/(1+i\omega/\omega_g)$, which has only a minor effect on $p$ even if the factor $1/(1+i\omega/\omega_g)$, a decreasing function of $\omega$, is replaced by unity  \cite{Iwasa2017}, the value at $\omega=0$.

In a special case of $m=K=0$, this equation turns into
\begin{align}
  (1+i \omega/\omega_\eta)(1+i\omega/\omega_g)p=\beta \overline P_\pm\cdot qv,
\end{align}
\noindent  low-pass with two time constants $\omega_g$ and $\omega_\eta$.

\section*{Results}

\begin{wrapfigure}{l}{0.65\textwidth} 
\vspace{-20pt} 
\includegraphics[width=0.66\textwidth]{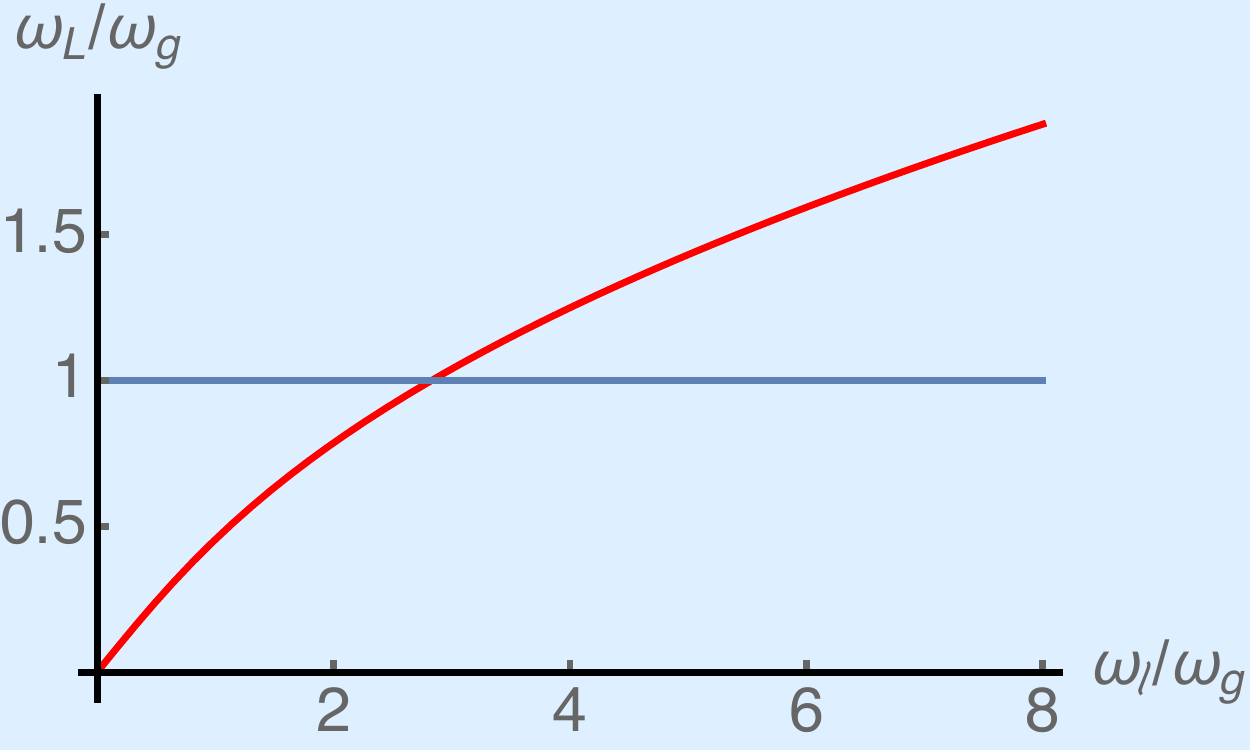}
\vspace{-20pt} 
\vspace{1pt}
\caption{Limiting frequency $\omega_L$ for prestin with gating frequency $\omega_g$ against that of $\omega_\ell$, for infinitely fast gating.}
\label{fig:freqL}
\end{wrapfigure}
The mechanical displacement of the OHC is given by $x=an p\cdot k/(k+K)$ and power output is $\eta\omega^2 |x|^2/2\pi$, i.e. proportional to $|p|^2$. Eq.\ \ref{eq:p1} shows that power output is attenuated from the value for infinitely fast gating by a factor $1/(1+\omega/\omega_g^2)$ by the speed of gating.

Recall how an optimal limiting frequency was determined if gating is infinitely fast. In the presence of inertia, power generation has a peak value $P_\infty^\mathrm{max}$ due to piezoelectric resonance.  A limiting frequency $\omega_\ell$ for that case is obtained from  
\[P_\infty^\mathrm{max}=\mu\omega_\ell^2, \]
equating the maximal power production with viscous loss. Here $\mu$ is proportional to viscous coefficient $\eta$.

If the resonance peak is sharp, the corresponding limiting frequency $\omega_{L}$ for a finite gating frequency $\omega_g$ can be approximated by
\begin{align}P_\infty^\mathrm{max}/[1+(\omega_L/\omega_g)^2] \approx \mu\omega_L^2. \end{align}

The combination of these two equations leads to
\bblubox{}
\begin{align}
(\omega_L/\omega_g)^2\approx\frac 1 2\left( \sqrt{1+4(\omega_\ell/\omega_g)^2}-1\right).
\end{align}
\eblubox
\noindent For $\omega_\ell=2\pi \times 10$ kHz \cite{Iwasa2017} and $\omega_g=2\pi\times 3$ kHz \cite{SantosSacchi2019}, $\omega_L/\omega_g=1.1$ (See Fig. \ref{fig:freqL}). The limiting frequency is 3.3 kHz, not much higher than the gating frequency.

\section*{Discussion}

\subsection*{Observations on the gating speed}
Experimental data, which constrain the gating speed of prestin, include the frequency dependence of quasi-isometric movement of OHCs, stimulated by capacitive coupling in the microchamber configuration \cite{fhg1999}, the frequency dependence of the membrane capacitance \cite{ga1997,dong2000} and current noise spectrum of sealed lateral membrane \cite{dong2000}, and whole-cell mode with capacitance compensation \cite{SantosSacchi2019}. 

Of these experimental observations, fast gating is consistent with the force measured under quasi-isometric condition using AFM cantilever, which does not roll-off up to 60 kHz \cite{fhg1999}. The current noise spectrum \cite{dong2000} suggests fast gating, above 40 kHz.

The most recent report \cite{SantosSacchi2019}, however, indicates that the intrinsic gating of prestin is about 3 kHz, much lower to cover the auditory range. 

The implication of the frequency dependence of the membrane capacitance of sealed lateral membrane of OHCs \cite{ga1997,dong2000} depends on the interpretation as to whether the limiting factor is the intrinsic gating or mechanical relaxation time.
\subsection*{Compatibility with the physiological role}
An upper bound of the limiting frequency estimated previously based on the fast gating model was about 10 kHz, consistent with the hearing range in the order of magnitude. If the gating frequency is \color{red} 3 kHz\color{black}, however, the corresponding limiting frequency is only \color{red} 3.3 kHz.\color{black} 


\section*{Conclusions}


\begin{itemize}
\item The power output of OHC can be significantly attenuated by the frequency of gating.
\item Gating frequency can dictate a limiting frequency of the local cochlear amplifier. 
\item A large gap between the gating frequency and the hearing frequency range is a big puzzle.
\end{itemize}


\color{black} 



\begin{thebibliography}{7}
\providecommand{\url}[1]{\texttt{#1}}
\providecommand{\urlprefix}{ }

\bibitem[Santos-Sacchi et~al.(2019)Santos-Sacchi, Iwasa, and
  Tan]{SantosSacchi2019}
Santos-Sacchi, J., K.~H. Iwasa, and W.~Tan, 2019.
\newblock Outer hair cell electromotility is low-pass filtered relative to the
  molecular conformational changes that produce nonlinear capacitance.
\newblock \emph{J Gen Physiol}

\bibitem[Iwasa(2016)]{Iwasa2016}
Iwasa, K.~H., 2016.
\newblock Energy Output from a Single Outer Hair Cell.
\newblock \emph{Biophys. J.} 111:2500--2511.

\bibitem[Iwasa(2017)]{Iwasa2017}
Iwasa, K.~H., 2017.
\newblock Negative membrane capacitance of outer hair cells: electromechanical
  coupling near resonance.
\newblock \emph{Sci. Rep.} 7:12118.

\bibitem[Iwasa(1997)]{i1997}
Iwasa, K.~H., 1997.
\newblock Current noise spectrum and capacitance due to the membrane motor of
  the outer hair cell: theory.
\newblock \emph{Biophys. J.} 73:2965--2971.

\bibitem[Frank et~al.(1999)Frank, Hemmert, and Gummer]{fhg1999}
Frank, G., W.~Hemmert, and A.~W. Gummer, 1999.
\newblock Limiting dynamics of high-frequency electromechanical transduction of
  outer hair cells.
\newblock \emph{Proc. Natl. Acad. Sci. USA} 96:4420--4425.

\bibitem[Gale and Ashmore(1997)]{ga1997}
Gale, J.~E., and J.~F. Ashmore, 1997.
\newblock An intrinsic frequency limit to the cochlear amplifier.
\newblock \emph{Nature} 389:63--66.

\bibitem[Dong et~al.(2000)Dong, Ehrenstein, and Iwasa]{dong2000}
Dong, X., D.~Ehrenstein, and K.~H. Iwasa, 2000.
\newblock Fluctuation of motor charge in the lateral membrane of the cochlear
  outer hair cell.
\newblock \emph{Biophys J} 79:1876--1882.

\end{thebibliography}

\end{document}